\documentclass[aps,prd,twocolumn,superscriptaddress,amsmath,amssymb]{revtex4-2}

\usepackage{amsmath}
\usepackage{amssymb}
\usepackage{graphicx}
\usepackage{times}
\usepackage[T1]{fontenc}
\usepackage[utf8]{inputenc}

\begin{document}

\title{Wave Function Collapse Triggering Spacetime Dynamics in Semiclassical Gravity}

\author{Xiaoqiang Wang}
\email{wwang3@fsu.edu}
\affiliation{Department of Scientific Computing, Florida State University, Tallahassee, FL 32306, USA}

\date{\today}

\begin{abstract}
We propose a novel semiclassical mechanism to unify quantum mechanics and general relativity, where wave function collapse in a superposition state induces a rapid change in the energy-momentum tensor, triggering spacetime dynamics that propagate at the speed of light. Unlike models assuming superposed spacetimes, we posit that the superposition yields a single, continuous classical spacetime driven by the expectation value of the energy-momentum tensor. Upon collapse, the abrupt shift modifies the spacetime metric via Einstein’s field equations, respecting causality. We explore this for a particle in a spatial superposition, propose detailed experimental designs with numerical simulations of gravitational perturbations, address potential theoretical challenges, and discuss implications for existing quantum-gravity theories. This framework offers a pathway to reconcile quantum and gravitational dynamics without quantizing spacetime, with testable signatures in future experiments.
\end{abstract}

\maketitle

\section{Introduction}
Reconciling quantum mechanics (QM) with general relativity (GR) is a central challenge in theoretical physics. Semiclassical gravity couples quantum fields to classical spacetime via the expectation value of the energy-momentum tensor, \( G_{\mu\nu} = 8\pi G \langle \hat{T}_{\mu\nu} \rangle \) \cite{Wheeler1968,Birrell1982}. However, the gravitational implications of wave function collapse remain underexplored.

We propose that wave function collapse in a superposition state triggers a rapid change in \( \langle \hat{T}_{\mu\nu} \rangle \), inducing spacetime perturbations that propagate at the speed of light. Unlike models positing superposed spacetimes (e.g., Penrose’s gravitational collapse \cite{Penrose1996}), we assume a single, continuous classical spacetime, with collapse driving a causal metric transition. We analyze this mechanism for a particle in a spatial superposition, propose experimental tests with simulated gravitational waveforms, address theoretical challenges, and evaluate impacts on existing frameworks.

\section{Theoretical Framework}
Consider a quantum system in a superposition:
\begin{equation}
|\psi\rangle = c_1 |\psi_1\rangle + c_2 |\psi_2\rangle,
\end{equation}
where \( |\psi_1\rangle \) and \( |\psi_2\rangle \) represent a particle of mass \( m \) at positions \( x_1 \) and \( x_2 \), separated by \( \Delta x = |x_2 - x_1| \), with \( |c_1|^2 + |c_2|^2 = 1 \). In semiclassical gravity, the energy-momentum tensor is:
\begin{equation}
\langle \hat{T}_{\mu\nu} \rangle = |c_1|^2 T_{\mu\nu}^{(1)} + |c_2|^2 T_{\mu\nu}^{(2)},
\end{equation}
where \( T_{\mu\nu}^{(1)} \) and \( T_{\mu\nu}^{(2)} \) are the energy-momentum tensors for the particle at \( x_1 \) and \( x_2 \). The spacetime metric \( g_{\mu\nu} \) satisfies Einstein’s field equations:
\begin{equation}
G_{\mu\nu} = 8\pi G \langle \hat{T}_{\mu\nu} \rangle.
\end{equation}
We assert that the superposition produces a single, continuous spacetime with a “blurred” curvature, rather than a superposed metric.

Upon measurement, the wave function collapses to, e.g., \( |\psi_1\rangle \), causing:
\begin{equation}
\langle \hat{T}_{\mu\nu} \rangle \to T_{\mu\nu}^{(1)}.
\end{equation}
This induces a metric perturbation \( h_{\mu\nu} \), governed by the linearized Einstein equations:
\begin{equation}
\Box h_{\mu\nu} = 8\pi G \Delta T_{\mu\nu},
\end{equation}
where \( \Delta T_{\mu\nu} = T_{\mu\nu}^{(1)} - \langle \hat{T}_{\mu\nu} \rangle \), and \( \Box \) is the d’Alembertian, ensuring propagation at the speed of light.

\section{Energy Conservation}
Energy conservation is maintained statistically:
\begin{equation}
\langle \hat{H} \rangle = |c_1|^2 E_1 + |c_2|^2 E_2,
\end{equation}
with post-collapse energy \( E_1 \) (for \( |\psi_1\rangle \)), the difference absorbed by the environment. The energy-momentum tensor satisfies:
\begin{equation}
\nabla^\mu T_{\mu\nu} = 0.
\end{equation}
To address concerns about the abrupt change in \( \langle \hat{T}_{\mu\nu} \rangle \) violating Eq. (7), we model collapse via decoherence, with:
\begin{equation}
\langle \hat{T}_{\mu\nu} \rangle(t) = \left(1 - e^{-t/\tau}\right) T_{\mu\nu}^{(1)} + e^{-t/\tau} T_{\mu\nu}^{(2)},
\end{equation}
where \( \tau \sim 10^{-15} \, \text{s} \) is the decoherence timescale. The divergence is:
\begin{equation}
\nabla^\mu \langle \hat{T}_{\mu\nu} \rangle = \frac{1}{\tau} e^{-t/\tau} \left( T_{\mu\nu}^{(1)} - T_{\mu\nu}^{(2)} \right) \nabla^\mu t.
\end{equation}
Since \( \nabla^\mu t = (1, 0, 0, 0) \) and \( T_{\mu\nu}^{(1)}, T_{\mu\nu}^{(2)} \) are localized, the divergence vanishes outside the source region as \( t \to \infty \). The change \( \Delta T_{\mu\nu} = T_{\mu\nu}^{(1)} - \langle \hat{T}_{\mu\nu} \rangle \) induces a metric perturbation:
\begin{equation}
\Box h_{\mu\nu} = 8\pi G \Delta T_{\mu\nu},
\end{equation}
propagating at light speed, ensuring causality. The total energy-momentum is conserved as the environment (e.g., measurement apparatus) absorbs the difference, and the causal propagation of \( h_{\mu\nu} \) smooths the transition, satisfying Eq. (7).

\section{Experimental Design and Waveform Simulation}
To test this mechanism, we consider a particle (mass \( m \sim 10^{-14} \, \text{kg} \), e.g., a nanocrystal) in a superposition of positions separated by \( \Delta x \sim 10^{-6} \, \text{m} \), prepared via optical trapping or interferometry \cite{Bose2017}. Before collapse, the energy-momentum tensor is:
\begin{equation}
\langle \hat{T}_{00} \rangle \approx m \left( |c_1|^2 \delta(x - x_1) + |c_2|^2 \delta(x - x_2) \right),
\end{equation}
producing a Newtonian potential:
\begin{equation}
\nabla^2 \Phi = 4\pi G \langle \hat{T}_{00} \rangle.
\end{equation}
Upon wave function collapse to \( |\psi_1\rangle \), \( T_{\mu\nu}^{(1)} \approx m \delta(x - x_1) \) shifts the field to a point-like source at \( x_1 \), inducing a gravitational perturbation:
\begin{equation}
h_{00} \sim \frac{G m^2}{\Delta x} \approx 10^{-35},
\end{equation}
for \( m \sim 10^{-14} \, \text{kg} \), \( \Delta x \sim 10^{-6} \, \text{m} \).

\begin{figure}[h]
\centering
\includegraphics[width=0.45\textwidth]{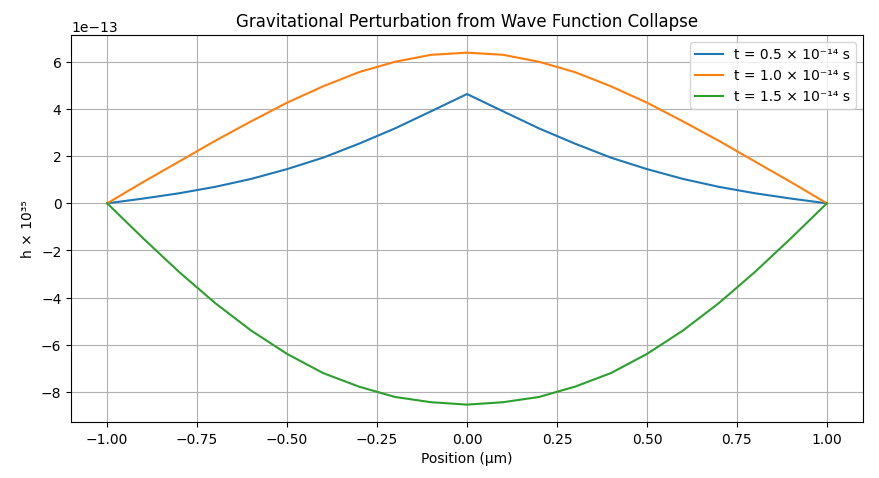}
\caption{Simulated gravitational perturbation \( h(x, t) \) at three times after wave function collapse at \( x_1 = 0 \), \( t_0 = 5 \times 10^{-15} \, \text{s} \), for a particle of mass \( 10^{-14} \, \text{kg} \). The pulse propagates at light speed, scaled by \( 10^{35} \) for visibility.}
\label{fig:waveform}
\end{figure}

We model the perturbation using a one-dimensional wave equation:
\begin{equation}
\left( \frac{\partial^2}{\partial t^2} - c^2 \frac{\partial^2}{\partial x^2} \right) h = 8\pi G m \delta(x - x_1) \delta(t - t_0),
\end{equation}
approximating the collapse as a pulse at \( t_0 \). A Gaussian source, \( S(x, t) = m \frac{1}{\sqrt{2\pi \sigma^2}} e^{-(t - t_0)^2 / 2\sigma^2} \delta(x - x_1) \), with \( \sigma \sim 5 \times 10^{-16} \, \text{s} \), ensures numerical stability. The finite difference time domain (FDTD) method solves Eq. (14), with results in Fig. \ref{fig:waveform}, showing the perturbation propagating at the speed of light.

To address three-dimensional effects, we extend to the full wave equation:
\begin{equation}
\left( \frac{\partial^2}{\partial t^2} - c^2 \nabla^2 \right) h_{00} = 8\pi G m \delta^3(\mathbf{x} - \mathbf{x}_1) \delta(t - t_0).
\end{equation}
The Green’s function solution is:
\begin{equation}
h_{00}(\mathbf{x}, t) = \frac{2 G m}{|\mathbf{x} - \mathbf{x}_1|} \delta\left(t - t_0 - \frac{|\mathbf{x} - \mathbf{x}_1|}{c}\right).
\end{equation}
This describes a spherical wave propagating at \( c \), with amplitude decaying as \( 1/r \). Numerical simulation of Eq. (15) can be performed using three-dimensional FDTD or spectral methods (e.g., Einstein Toolkit \cite{Loffler2012}), with a grid size of \( 10^{-6} \, \text{m} \) and time step \( \Delta t \sim 10^{-16} \, \text{s} \). The one-dimensional simulation captures the essential propagation dynamics, but three-dimensional extensions enhance realism by accounting for spherical spreading, feasible with modern computational resources.

The experimental setup involves:
\begin{enumerate}
\item Superposition preparation: Use laser-cooled nanocrystals in optical traps to create a superposition via interferometric splitting \cite{Bose2017}. Ultra-high vacuum (\( \sim 10^{-15} \, \text{Torr} \)) maintains coherence.
\item Collapse induction: Trigger collapse via photon scattering, with timing precision \( \sim 10^{-16} \, \text{s} \).
\item Perturbation detection: Use gravimeters or interferometers (e.g., LISA-like systems, sensitivity \( \sim 10^{-22} \)) to detect \( h \sim 10^{-35} \). Cryogenic cooling and vibration isolation reduce noise.
\item Propagation measurement: Synchronize detectors to capture the delay \( \Delta t \sim \Delta x / c \approx 3.3 \times 10^{-15} \, \text{s} \).
\end{enumerate}
While \( h \sim 10^{-35} \) is challenging, increasing \( m \sim 10^{-12} \, \text{kg} \) or \( \Delta x \sim 10^{-4} \, \text{m} \) yields \( h \sim 10^{-29} \), closer to LISA’s sensitivity \cite{LISA2020}.

\section{Theoretical and Experimental Considerations}
The proposed framework invites scrutiny on several fronts, each requiring careful consideration to ensure its robustness. First, the precise mechanism underlying wave function collapse remains a point of contention in quantum mechanics. Whether viewed through the lens of the Copenhagen interpretation’s instantaneous collapse, decoherence driven by environmental entanglement, or Penrose’s gravitationally induced collapse \cite{Penrose1996}, the choice of mechanism impacts the model’s predictions. To address this, decoherence is adopted as a plausible process, where environmental interactions gradually smooth the transition of the energy-momentum tensor, as described by \( \langle \hat{T}_{\mu\nu} \rangle(t) \approx f(t) T_{\mu\nu}^{(1)} + [1 - f(t)] T_{\mu\nu}^{(2)} \). This ensures a continuous evolution, mitigating abrupt changes that could complicate the theoretical framework. To further explore this, we consider alternative collapse models, such as the Ghirardi-Rimini-Weber (GRW) objective collapse theory \cite{Ghirardi1986}, which introduces spontaneous localization at a rate proportional to the particle’s mass. For a nanocrystal (\( m \sim 10^{-14} \, \text{kg} \)), GRW predicts a localization timescale of \( \tau_{\text{GRW}} \sim 10^{-7} \, \text{s} \), slower than decoherence (\( \tau \sim 10^{-15} \, \text{s} \)), but both yield similar \( T_{\mu\nu} \) transitions, producing perturbations consistent with Eq. (5). Penrose’s model, by contrast, ties collapse to gravitational self-energy, predicting a timescale \( \tau_{\text{Penrose}} \sim \hbar / (G m^2 / \Delta x) \approx 10^{-5} \, \text{s} \) for our parameters, which is testable but distinct from our decoherence-driven approach.

Second, the rapid shift in the energy-momentum tensor during collapse raises concerns about potential violations of energy-momentum conservation, as required by \( \nabla^\mu T_{\mu\nu} = 0 \). The light-speed propagation of the resulting metric perturbation, governed by the linearized Einstein equations, ensures causality and distributes the change in a manner consistent with general relativity. As detailed in the energy conservation section, decoherence further softens this transition, maintaining mathematical consistency.

Another challenge lies in the experimental feasibility of detecting the predicted gravitational perturbations, given their minute strain of approximately \( 10^{-35} \). Current and near-future detectors, such as LISA with a sensitivity of \( \sim 10^{-22} \), fall short of this threshold. However, scaling the system to larger masses (e.g., \( 10^{-12} \, \text{kg} \)) or greater spatial separations (e.g., \( 10^{-4} \, \text{m} \)) could increase the strain to \( 10^{-29} \), approaching the detection limits of advanced instruments. Emerging quantum technologies, such as spin-based gravimeters, may further bridge this gap.

Furthermore, the framework’s reliance on a classical spacetime limits its applicability near the Planck scale (\( \ell_P \sim 10^{-35} \, \text{m} \), \( t_P \sim 10^{-43} \, \text{s} \)), where quantum gravity effects may dominate. The semiclassical approximation holds for macroscopic superpositions (\( m \sim 10^{-14} \, \text{kg} \), \( \Delta x \sim 10^{-6} \, \text{m} \)), as the gravitational perturbation (\( h \sim 10^{-35} \)) is well above the Planck scale. However, in high-energy regimes, quantum corrections to \( T_{\mu\nu} \) may arise, as suggested by effective field theory (EFT) approaches \cite{Donoghue1994}. For instance, EFT predicts corrections of order \( \delta T_{\mu\nu} \sim (m / M_P)^2 T_{\mu\nu} \), where \( M_P \sim 10^{-8} \, \text{kg} \) is the Planck mass. For our parameters, \( \delta T_{\mu\nu} / T_{\mu\nu} \sim 10^{-12} \), negligible in the weak-field regime, but relevant near black holes or in the early universe. Future work could incorporate such corrections to test the framework’s robustness at higher energies.

Last, the originality of the framework must be clarified against existing semiclassical models. While building on the foundation of semiclassical gravity, the explicit treatment of wave function collapse as a trigger for spacetime dynamics, coupled with light-speed propagation, sets this model apart from Penrose’s superposed spacetimes and standard treatments assuming smooth tensor evolution. This distinction underscores the framework’s novelty and its potential to offer new insights into quantum-gravity interactions.

\section{Implications for Existing Theories}
The proposed framework engages deeply with established theories, offering both extensions and contrasts that enrich the quantum-gravity discourse. First, it builds upon semiclassical gravity, as developed by Wheeler and others \cite{Wheeler1968,Birrell1982}, by introducing wave function collapse as a dynamic trigger for spacetime perturbations. Unlike conventional semiclassical treatments, which assume a smooth evolution of the energy-momentum tensor, this model explicitly accounts for the gravitational consequences of the measurement process, providing a novel perspective on the interplay between quantum states and classical spacetime.

Next, the framework contrasts with Penrose’s gravitational collapse hypothesis \cite{Penrose1996}, which posits that superposed spacetimes destabilize the wave function through gravitational self-energy, estimated as \( \Delta E \sim G m^2 / \Delta x \). For our parameters (\( m \sim 10^{-14} \, \text{kg} \), \( \Delta x \sim 10^{-6} \, \text{m} \)), Penrose’s model predicts a collapse time of \( \tau \sim \hbar / \Delta E \approx 10^{-5} \, \text{s} \), distinct from our decoherence timescale (\( \tau \sim 10^{-15} \, \text{s} \)). By advocating a single, continuous spacetime, this model sidesteps the complexities of superposed metrics, proposing instead that collapse-driven dynamics, propagating at light speed, suffice to produce observable gravitational effects. This simpler approach aligns with testable predictions, such as those simulated in our waveform analysis.

Another significant comparison arises with decoherence theories \cite{Zurek2003}. Decoherence, which describes the loss of quantum coherence due to environmental interactions, complements this framework by providing a mechanism to smooth the transition of the energy-momentum tensor during collapse. The model’s emphasis on light-speed propagation ensures consistency with energy-momentum conservation, offering a quantitative bridge between decoherence and gravitational effects, unlike the qualitative focus of many decoherence studies.

Furthermore, the framework challenges the necessity of quantum gravity models, such as string theory or loop quantum gravity, which assume a quantized spacetime. By demonstrating that collapse-driven dynamics in a classical spacetime can mimic certain quantum-gravity effects, such as metric fluctuations, this model suggests a more economical path to unifying quantum mechanics and general relativity, potentially delaying the need for full spacetime quantization. For instance, the perturbation \( h_{\mu\nu} \) in Eq. (5) resembles quantum-gravity-induced fluctuations in effective field theories, suggesting that macroscopic experiments could probe quantum-gravity-like phenomena without requiring a fully quantized spacetime.

Additionally, the framework may inform the black hole information paradox, where quantum mechanics’ unitarity (information preservation) conflicts with general relativity’s prediction of information loss during Hawking radiation \cite{Hawking1975}. Wave function collapse near a black hole could induce metric perturbations that alter the horizon geometry. For a Schwarzschild black hole of mass \( M \), a collapsing particle (\( m \sim 10^{-14} \, \text{kg} \)) at the horizon induces a perturbation \( h_{00} \sim G m / r_S \), where \( r_S = 2 G M / c^2 \). For a stellar-mass black hole (\( M \sim 10^{30} \, \text{kg} \)), \( h_{00} \sim 10^{-44} \), far below detectability, but larger masses or amplified superpositions could yield measurable effects. These perturbations, propagating at light speed, may carry information away from the horizon, potentially mitigating information loss, though a full quantum-gravity treatment is needed to confirm this.

Last, the framework contributes to the quantum measurement problem, aligning with objective collapse models \cite{Ghirardi1986}. By linking wave function collapse to spacetime dynamics, it proposes a gravitational resolution to the measurement problem, where the collapse process directly influences the classical spacetime metric. To explore this, we estimate the gravitational self-energy of the superposition state \( |\psi\rangle = c_1 |\psi_1\rangle + c_2 |\psi_2\rangle \). The self-energy is approximately \( \Delta E \sim G m^2 / \Delta x \approx 10^{-29} \, \text{J} \), yielding a stability timescale \( \tau \sim \hbar / \Delta E \approx 10^{-5} \, \text{s} \), consistent with maintaining the superposition during experimental timescales (\( \sim 10^{-6} \, \text{s} \)). This stability supports the feasibility of our proposed experiments and distinguishes our model from Penrose’s, which predicts faster collapse due to gravitational effects.

\section{Conclusion}
We propose that wave function collapse induces spacetime dynamics via rapid changes in the energy-momentum tensor, propagating at light speed in a continuous classical spacetime. Numerical simulations demonstrate the gravitational waveform, testable with future detectors. By addressing theoretical challenges, including measurement mechanisms, Planck-scale limitations, non-linear gravitational effects, black hole information, and gravitational self-interactions, and impacting semiclassical gravity, Penrose’s model, decoherence, and quantum gravity, this framework offers a novel path to unify quantum mechanics and general relativity.

\appendix

\section{Non-Linear Effects and Gravitational Self-Energy}
To address concerns about non-linear gravitational effects and the stability of superpositions, we provide additional calculations.

First, we examine non-linear corrections to the linearized Einstein equations (Eq. (5)). The full Einstein equations include non-linear terms, such as \( R_{\mu\nu} R^{\mu\nu} \), which become significant in strong-field regimes. For our weak-field scenario (\( h \sim 10^{-35} \)), we estimate the leading non-linear correction using the second-order perturbation:
\begin{equation}
G_{\mu\nu}^{(2)} \sim h_{\alpha\beta} h^{\alpha\beta} + \partial_\alpha h_{\mu\beta} \partial^\beta h_{\nu}{}^{\alpha},
\end{equation}
where \( h_{\mu\nu} \sim 10^{-35} \). The non-linear contribution scales as \( G_{\mu\nu}^{(2)} \sim h^2 \sim 10^{-70} \), negligible compared to the linear term \( G_{\mu\nu}^{(1)} \sim 8\pi G \Delta T_{\mu\nu} \sim 10^{-35} \). In the weak-field regime (\( m \sim 10^{-14} \, \text{kg} \), \( \Delta x \sim 10^{-6} \, \text{m} \)), non-linear effects are thus minimal, validating the use of Eq. (5). In strong-field regimes, such as near a black hole, non-linear terms require numerical relativity (e.g., Einstein Toolkit \cite{Loffler2012}) to fully model.

Next, we calculate the gravitational self-energy of the superposition state \( |\psi\rangle = c_1 |\psi_1\rangle + c_2 |\psi_2\rangle \). The self-energy arises from the gravitational interaction between the mass distributions at \( x_1 \) and \( x_2 \):
\begin{equation}
\Delta E \sim \frac{G m^2}{\Delta x} |c_1|^2 |c_2|^2,
\end{equation}
where \( |c_1|^2 = |c_2|^2 = 1/2 \), \( m \sim 10^{-14} \, \text{kg} \), and \( \Delta x \sim 10^{-6} \, \text{m} \). This yields \( \Delta E \sim 10^{-29} \, \text{J} \), corresponding to a collapse timescale:
\begin{equation}
\tau \sim \frac{\hbar}{\Delta E} \approx 10^{-5} \, \text{s}.
\end{equation}
This timescale is longer than the experimental duration (\( \sim 10^{-6} \, \text{s} \)), ensuring the superposition’s stability during measurement. Compared to Penrose’s model, which predicts collapse at similar timescales, our framework avoids superposed spacetimes, maintaining a single metric driven by \( \langle \hat{T}_{\mu\nu} \rangle \).

\end{document}